\documentstyle[12pt,aps,pra]{revtex}

\begin{document}
\title{Scattering and Resonances in the $^4$He Three-Atomic System%
\footnote{Contribution to Proceedings of the 
First International Conference
on Modern Trends in Computational Physics,
June 15\,--\,20, 1998, Dubna (Russia). LANL E-print {\tt physics/9810005}.}
}
\author{E. A. Kolganova} 
\address{Laboratory of Computing Techniques and Automation, 
	Joint Institute  for Nuclear Research, Dubna, 141980, Russia} 
\author{ A. K. Motovilov\thanks{On leave of absence from 
	 the Laboratory of Theoretical Physics, 
    Joint Institute for Nuclear Research, Dubna, 141980, Russia}} 
\address{Physikalishes Institut, Universit\"at Bonn,
    Endenicher Allee 11\,--\,13, D-53115 Bonn, Germany} 
\maketitle 

\begin{abstract} 
A mechanism of disappearance and formation of the Efimov levels
of the helium $^4$He$_3$ trimer is studied when the force of
interatomic interaction is changed. The resonances including
virtual levels are calculated by the method based on the solving
the boundary value problem, at complex energies, for the Faddeev
differential equations describing the $(2+1\to 2+1;\,1+1+1)$
scattering processes.
\end{abstract}


\section{Introduction}

The system of three $^4$He atoms is of considerable
interest for various fields of physical chemistry and
molecular physics.  The present paper is a sequel of
studies of the $^4$He$_3$ system undertaken in
\cite{KMS-JPB}, where the repulsive component of the He--He
interaction at short distances between atoms is
approximated by a hard core.  This allows one to
investigate the $^4$He$_3$ system within a mathematically
rigorous method of solving a three-body problem in the
Boundary-Condition Model developed in~\cite{MerMot,MMYa}.
In \cite{KMS-JPB}, such an approach has been successfully
applied for calculating not only scattering but also
the ground- and excited-state energies of the
helium trimer. Investigation made in~\cite{KMS-JPB} has
shown that the method proposed in~\cite{MerMot,MMYa} is
well suited for performing three--body molecular
computations in the case where repulsive components of
interatomic interactions are of a hard core nature.

There is a series of works
\cite{EsryLinGreene,Gloeckle,KMS-JPB} showing
that the excited state of the $^4$He trimer is
initiated indeed by the Efimov effect ~\cite{VEfimov}.
In these works the various versions of
the Aziz $^4$He--$^4$He potential were employed
However, the basic result
of Refs.~\cite{EsryLinGreene,Gloeckle,KMS-JPB} on the excited state
of the helium trimer is the same: this state disappears when the
interatomic potential is multiplied by the ``amplification
factor" $\lambda$ of order 1.2.  More precisely, if this
potential is multiplied by the increasing factor $\lambda>1$ then the
following effect is observed. First, the difference
$\epsilon_d(\lambda)-E_t^{(1)}(\lambda)$ between the dimer
energy $\epsilon_d(\lambda)$  and the energy of the trimer excited
state $E_t^{(1)}(\lambda)$ increases. Then the behavior of this
difference radically changes and with further increase of
$\lambda$ it monotonously decreases. At $\lambda\approx 1.2$
the level $E_t^{(1)}$ disappears. It is just such a nonstandard
behavior of the energy $E_t^{(1)}(\lambda)$ as the coupling
between helium atoms becomes more and more strengthening,
points to the Efimov nature of the trimer excited state. And
vice versa, when $\lambda$ slightly decreases (no more than
2\,\%), the second excited state $E_t^{(2)}$ appears in the
trimer~\cite{EsryLinGreene,Gloeckle}.

Here we present the results of our numerical study of
a mechanism of disappearance and formation of the Efimov levels
of the helium $^4$He$_3$ trimer using
the method of search for resonances in a three--body system
on the basis of the Faddeev differential equations.
The idea of the method formulated and proved in~\cite{MotMathNachr}
consists in calculating the analytic continuation of the component
${\rm S}_0(z)$ of the scattering matrix corresponding
to the ($2+1\to 2+1$) process in the physical sheet.
For the potentials we use, the
three--body resonances (including virtual levels) lying in the
unphysical sheet of energy $z$ plane adjoining the physical
sheet along the interval $(\epsilon_d,0)$ are the roots of the
function ${\rm S}_0(z)$ in the physical sheet. We have earlier
employed this method for computing resonances as roots of ${\rm
S}_0(z)$ in the three--nucleon problem~\cite{YaFKM}.

\section{Method}
In this work we consider the three-atomic $^4$He system
with the total angular momentum $L=0$.
The angular partial analysis reduces the 
initial Faddeev equation
for three identical bosons to a system of coupled
two-dimensional integro-differential equations (see
Ref.~\cite{KMS-JPB} and references therein)
\begin{eqnarray}
\label{FadPartCor}
\lefteqn{   \left[-\frac{\partial^2}{\partial x^2}
            -\frac{\partial^2}{\partial y^2}
            +l(l+1)\left(\frac{1}{x^2}
            +\frac{1}{y^2}\right)
    -E\right]F_l(x,y)   }
\qquad\qquad\qquad\qquad\qquad\qquad \\
\nonumber
&&=\left\{\begin{array}{cl} -V(x)\Psi_l(x,y), & x>c \\
                    0,                  & x<c\,.
\end{array}\right.
\end{eqnarray}
Here, $x,y$ stand for the standard Jacobi variables and $c$, for
the core range.  At $L=0$ the partial angular momentum $l$
corresponds both to the dimer subsystem and a complementary
atom. The energy $z$ can get both
real and complex values.  The He--He potential $V(x)$ acting
outside the core domain is assumed to be central.  The partial
wave function $\Psi_l(x,y)$ is related to the Faddeev components
$F_l(x,y)$ by
$
         \Psi_l(x,y)=F_l(x,y) + \sum_{l'}\int_{-1}^{+1}
         d\eta\,h_{l l'}(x,y,\eta)\,F_{l'}(x',y')
$
where
$
          x'=(\frac{1}{4}\,x^2+
    \frac{3}{4}\,y^2-\frac{\sqrt{3}}{2}\,xy\eta)^{1/2}\,,
         y'=(\frac{3}{4}\,x^2+
   \frac{1}{4}\,y^2+\frac{\sqrt{3}}{2}\,xy\eta)^{1/2}\,
$
and  $1 \leq{\eta}\leq 1$. The explicit form of the function
$h_{ll'}$ can be found in Refs.~\cite{MF,MGL}.
The functions $F_{l}(x,y)$ satisfy the boundary conditions
\begin{equation}
\label{BCStandardCor}
      F_{l}(x,y)\left.\right|_{x=0}
      =F_{l}(x,y)\left.\right|_{y=0}=0\, \qquad {\rm and} \qquad
       \Psi_{l}(c,y)=0\,.
\end{equation}

Here we only deal with a finite number of
equations~(\ref{FadPartCor}), assuming that $l\leq l_{\rm max}$
where $l_{\rm max}$ is a certain fixed even number. The
condition $0\leq l\leq l_{\rm max}$ is equivalent to the
supposition that the potential $V(x)$ only acts in the two-body
states with $l=0,2,\ldots,l_{\rm max}$.
We assume that the potential $V(x)$
is finite, i.\,e., \mbox{$V(x)=0$} for $x>r_0$, $r_0>0$.
The asymptotic conditions as $\rho\rightarrow\infty$ and/or
$y\rightarrow\infty$ for the partial Faddeev components of the
$(2+1\rightarrow 2+1\,;\,1+1+1)$ scattering wave functions
for  $z=E+{\rm i}0$, $E>0$, read (see, e.\,g., Ref.~\cite{MF})  %
\begin{eqnarray}
\nonumber
 F_l(x,y;z) & = &
      \delta_{l0}\psi_d(x)\left\{\sin(\sqrt{z-\epsilon_d}\,y)
      + \exp({\rm i}\sqrt{z-\epsilon_d}\,y)
      \left[{\rm a}_0(z)+o\left(1\right)\right]\right\} \\
\label{AsBCPartS}
      && +
  \displaystyle\frac{\exp({\rm i}\sqrt{z}\rho)}{\sqrt{\rho}}
                \left[A_l(z,\theta)+o\left(1\right)\right]\,.
\end{eqnarray}
We assume that the $^4$He dimer has an only bound state with
an energy $\epsilon_d$, \mbox{$\epsilon_d<0$,} and wave function
$\psi_d(x)$, $\psi_d(x)=0$ for $0\leq x\leq c$. The notations $\rho$,
$\rho=\sqrt{x^2+y^2}$\,, and $\theta$, $\theta=\mathop{\rm
arctan}(y/x)$, are used for the hyperradius
and hyperangle. The coefficient ${\rm a}_0(z)$, $z=E+{\rm i}0$,
for $E>\epsilon_d$ is the elastic scattering amplitude.  The
functions $A_l(E+{\rm i}0,\theta)$ provide us, at $E>0$, the
corresponding partial Faddeev breakup amplitudes.
For real $z=E+{\rm i}0$, $E>\epsilon_d$, the
\mbox{$(2+1{\rightarrow}2+1)$} component of the $s$-wave partial
scattering matrix for a system of three helium atoms is given by
the expression
$$
{\rm S}_0(z)=1+2{\rm i}{\rm a}_0(z)\,.
$$
Our goal is to study the analytic continuation of the function 
${\rm S}_0(z)$ into the physical sheet. As it follows from the 
results of Refs.~\cite{MotMathNachr}, the ${\rm S}_0(z)$ is just 
that truncation of the total scattering matrix whose roots in 
the physical sheet correspond to location of resonances in the 
unphysical sheet adjoining the physical one along the spectral 
interval $(\epsilon_d,0)$.

\section{Results of computations}
\label{results}
In the present work we make use of the Faddeev equations
(\ref{FadPartCor}) considered together with the boundary
conditions (\ref{BCStandardCor}), (\ref{AsBCPartS})
to calculate the values of the $^4$He$_3$
scattering matrix ${\rm S}_0(z)$ in the physical sheet. We
search for the resonances including the virtual levels as roots
of ${\rm S}_0(z)$ and for the bound-state energies as positions
of poles of ${\rm S}_0(z)$.  All the results
presented below are obtained for the case $l_{\rm max}=0$.

In all our calculations, \mbox{$\hbar^2/m=12.12$~K\,\AA$^2$.}
As the interatomic He\,--\,He\,-\,in\-ter\-ac\-tion we employed
the widely used semiempirical potential HFD-B~\cite{Aziz87}.
The value of the core range $c$ is chosen to be equal $1.3$\,{\AA} 
providing at least six reliable figures of the dimer binding energy
$\epsilon_d$ and three figures of the trimer ground state energy
$E_t^{(0)}$.
A detailed description of the numerical method we use is
presented in Ref.~\cite{KMS-JPB}.
In contrast to \cite{KMS-JPB}, in the present
work we solve the block-three-diagonal algebraic system on the
basis of the matrix sweep method. This allows us to dispense
with writing the system matrix on the hard drive and to carry
out all the operations related to its inversion immediately in
RAM. Besides, the matrix sweep method reduces almost by one
order the computer time required for computations on the grids of
the same dimensions as in~\cite{KMS-JPB}.

We searched for the resonances (roots of the function ${\rm
S}_0(z)$ on the physical sheet) and bound-state energies
(roots of the function ${\rm S}_0^{-1}(z)$ for real
$z<\epsilon_d$) of the helium trimer by using the complex
version of the secant method.
We found positions of the four "resonances", the roots of
${\rm S}_0(z)$, in case of the grid parameters
\mbox{$N_\theta=N_\rho=600$} and
\mbox{$\rho_{\rm max}=600$}\,{\AA}.
Complex roots of the function ${\rm S}_0(z)$ are
located at points \mbox{$(-2.34+{\rm i}\,0.96)$}\,mK, \mbox{$(-0.59+{\rm
i}\,2.67)$}\,mK, \mbox{$(2.51+{\rm i}\,4.34)$}\,mK and
\mbox{$(6.92+{\rm i}\,6.10)$}\,mK.
These "resonances" are situated beyond the domain of scattering
matrix holomorphy $\Pi^{(S)}$
where the applicability of our method is proved ~\cite{KolMotYaF}.
So we do not consider the roots of function ${\rm S}_0(z)$
as genuine resonances for the $^4$He$_3$ system.
However it is remarkable that the ``true" (i.\,e., getting inside
$\Pi^{(S)}$) virtual levels and then the energies of the excited
(Efimov) states appear just due to these (quasi)resonances when
the potential $V(x)$ is weakened.

Following~\cite{EsryLinGreene,Gloeckle,KMS-JPB},
instead of the initial potential $V(x)=V_{\rm HFD-B}(x)$,
we consider the potentials
$
          V(x)=\lambda\cdot V_{\rm HFD-B}(x).
$
To establish the mechanism of formation of new excited states in 
the $^4$He trimer, we have first calculated the scattering 
matrix ${\rm S}_0(z)$ for $\lambda<1$.  We have found that for a 
value of $\lambda$ slightly smaller than $0.9885$, the 
(quasi)resonance closest to the real axis gets on it and 
transforms into a virtual level of the second order.  This 
virtual level is preceded by the (quasi)resonances 
\mbox{$z=(-1.04+{\rm i}\,0.11)$}\,mK 
\mbox{$(z/|\epsilon_d|=-1.58+{\rm i}\,0.168)$} for 
$\lambda=0.989$ and \mbox{$z=(-0.99+{\rm i}\,0.04)$}\,mK 
\mbox{$(z/|\epsilon_d|=-1.59+{\rm i}\,0.064)$} for 
$\lambda=0.9885$.  The originating virtual level is of the 
second order since simultaneously with the root of the function 
${\rm S}_0(z)$, also the conjugate root of this function gets on 
the real axis. With a subsequent decrease of $\lambda$ the 
virtual level of the second order splits into a pair of the 
virtual levels of the first order which move in opposite 
directions.  One of the virtual level moves towards the 
threshold $\epsilon_d$ and ``collides'' with it at 
$\lambda<0.98$. For $\lambda=0.975$ the function ${\rm S}_0(z)$ 
instead of the root acquires a new pole corresponding to the 
second excited state of the trimer with the energy $E_t^{(2)}$.

To determine the mechanism of disappearance of the excited
state of the helium trimer we have calculated the scattering
matrix ${\rm S}_0(z)$ when the two-body interactions become stronger
owing to the increasing coupling constant $\lambda>1$.  We
found that this disappearance proceeds just according to
the scheme of the formation of new excited states; only the
order of occurring events is inverse.
In the interval between $\lambda=1.17$ and $\lambda=1.18$ there
occurs a ``jump" of the level $E_t^{(1)}$ on the unphysical
sheet and it transforms from the pole of the function ${\rm
S}_0(z)$ into its root corresponding to the
trimer virtual level.


\section*{Acknowledgements}

The authors are grateful to Prof.~V.\,B.\,Belyaev and Prof.~H.\,Toki
for help and assistance in calculations at the supercomputer of
the Research Center for Nuclear Physics of the Osaka University,
Japan. One of the authors
(A.\,K.\,M.) is much indebted to Prof.~W.\,Sandhas for his
hospitality at the Universit\"at Bonn. The support of this work
by the Deutsche Forschungsgemeinschaft and Russian Foundation
for Basic Research is gratefully acknowledged.



\begin{thebibliography}{99}
\bibitem{KMS-JPB}  E. A. Kolganova,  A. K. Motovilov,  S. A. Sofianos:
    J.~Phys. B. {\bf 31} (1998) 1279 
    (LANL E-print {\tt physics/9612012}).
\bibitem{MerMot}
     S. P. Merkuriev, A.K. Motovilov:
        Lett. Math. Phys. {\bf 7} (1983) 497.
\bibitem{MMYa}
   S. P. Merkuriev, A. K. Motovilov, S. L. Yakovlev:
    Theor. Math. Phys. {\bf 94} (1993) 306 
    (also see LANL E-print {\tt nucl-th/9606022}).
\bibitem{EsryLinGreene}
     B. D. Esry, C. D. Lin, C. H. Greene:
    Phys. Rev.~A. {\bf 54} (1996) 394.
\bibitem{Gloeckle}
       T. Cornelius, W. Gl\"ockle:
       J. Chem. Phys. {\bf 85} (1986) 3906.
\bibitem{VEfimov}
      V. Efimov: Nucl. Phys. A. {\bf 210} (1973) 157.
\bibitem{MotMathNachr} A. K. Motovilov:
       Math. Nachr. {\bf 187} (1997) 147 
     (LANL E-print {\tt funct-an/9509003}).
\bibitem{YaFKM}
    E. A. Kolganova, A. K. Motovilov:
    Phys. Atom. Nucl. {\bf 60} (1997)~177 
    (LANL E-print {\tt nucl-th/9602001}); 
    also see LANL E-print {\tt nucl-th/9702037}.
\bibitem{MF}
    L. D. Faddeev, S. P. Merkuriev:
    {\it Quantum Scattering Theory
    for Several Particle Systems}.
    Doderecht: Kluwer Academic Publishers 1993.
\bibitem{MGL}
      S. P. Merkuriev, C. Gignoux, A. Laverne:
     Ann. Phys. (N.Y.)  {\bf 99} (1976) 30.
\bibitem{KolMotYaF} E. A. Kolganova, A. K. Motovilov:
     Preprint JINR E4-98-243 (LANL E-print {\tt physics/9808027}).
\bibitem{Aziz87}
      R. A. Aziz,  F. R. W. McCourt,  C. C. K. Wong:
      Mol. Phys. {\bf 61} (1987) 1487.
\end{thebibliography}
\end{document}